\documentclass[a4paper,useAMS,usenatbib,fleqn]{mnras}
\usepackage{graphics,graphicx,times}
\usepackage{amsmath}
\usepackage{amssymb}
\usepackage{aas_macros}
\usepackage[usenames,dvipsnames]{xcolor}
\usepackage{mathptmx}

\newcommand{\eagle}{{\sc eagle}}

\title[Origin of subgroups of CEMP(-no)]{The chemical imprint of the bursty nature of Milky Way's progenitors}
\author[Mahavir Sharma et al.]
{Mahavir Sharma\thanks{mahavir.sharma@durham.ac.uk}, Tom Theuns \& Carlos Frenk\\
Institute for Computational Cosmology, Department of Physics, University of Durham, South Road, Durham, DH1 3LE, UK\\
}

\begin{document}

\date{Submitted ---------- ; Accepted ----------; In original form ----------}


\maketitle
\begin{abstract}
Carbon enhanced metal poor (CEMP) stars with low abundances of neutron capture elements (CEMP-no stars) are ubiquitous among metal poor stars in the Milky Way. Recent observations have uncovered their two subgroups that differ in the carbon to magnesium ([C/Mg]) abundance ratio. Here we demonstrate that similar abundance patterns are also present in Milky Way-like galaxies in the \eagle\ cosmological hydrodynamical simulation, where these patterns originate from the fact that stars may form from gas enriched predominantly by asymptotic giant branch (AGB) stars or by type-II supernovae. This occurs when stars form in the poorly mixed interstellar medium of Milky Way progenitor galaxies that are undergoing bursty star formation. The CEMP-no stars with lower [C/Mg] form at the onset of a starburst from gas enriched by low metallicity type-II supernovae that power a strong outflow, quenching further star formation. When star formation resumes following cosmological gas accretion, the CEMP-no stars with higher [C/Mg] form, with enrichment by AGB ejecta evident in their higher abundance of barium and lower abundance of magnesium. This suggests that bursty star formation in the progenitors of the Galaxy leaves a permanent imprint in the abundance patterns of CEMP
 	stars.
\end{abstract}
\section{Introduction}
The quest for finding the most primitive generation of stars that formed in the Universe has led to the discovery of a population of low metallicity Milky Way (MW) stars that are enhanced in carbon relative to iron (termed CEMP stars, \citealt{Beers05}). In fact, the majority of the lowest metallicity stars currently found in the Milky Way and its dwarf satellites are CEMP stars \citep{Frebel15,Yoon16}. It has been suggested that their peculiar abundance pattern results from enrichment by supernovae from the first stars \citep[e.g.][]{Umeda03,Umeda05,Frebel08,Nomoto13,Keller14,Tominaga14} or by pair instability supernovae \citep{Aoki14,Jeon17}. Recently, \citet{Sharma17} argued that CEMP stars were enriched by, and formed contemporaneously with, the massive stars that reionized the Universe (\lq reionizers\rq), making CEMP stars the \lq siblings of reionizers\rq\ (SoRs). The abundance pattern of CEMP stars, specifically the lack of iron compared to carbon at low metallicity has recently become a focus of much theoretical research \citep[e.g.][]{Chiaki17,Sharma18}, and it has been argued that abundance patterns of CEMP stars are governed by the physical processes that operate in the interstellar medium of proto-galaxies, and hence they provide a window to the era when the first galaxies were being formed and the Universe was going through a transition from neutral to ionized \citep[e.g.][]{Chiappini13,Sharma17}.

In the classification of \citet{Yoon16}, CEMP stars have\footnote{We use the common notation [X/Y] to denote the logarithmic ratio of number density of element X over element Y in a star compared to that ratio in the Sun, [X/Y]$\equiv \log({\rm N}_{\rm X}/{\rm N}_{\rm Y}) - \log({\rm N}_{\rm X}/{\rm N}_{\rm Y})_\odot$.}  [Fe/H]$<-1$ and [C/Fe]$>0.7$. They are divided in subclasses based on their abundance of neutron capture elements (\citealt{Beers05}, see also \citealt{Masseron10}): CEMP-s stars (rich in s-process - slow neutron capture - elements such as barium), CEMP-r stars (rich in r-process - rapid neutron capture - elements, such as europium), rich in both types of neutron capture elements (CEMP-r/s) or  not overabundant in either neutron capture element (CEMP-no stars). \citet{Aoki07} found that the CEMP stars exhibit a bimodal distribution in [Fe/H], with most of CEMP-no occupying the extremely metal poor regime ([Fe/H]$<-3$) and CEMP-s having a relatively higher amount of [Fe/H] $\sim-2$. More recent studies demonstrate that the CEMP-no fraction increases with decreasing [Fe/H], such that at [Fe/H]$<-4$ most are CEMP-no \citep{Placco14,Frebel15,Yoon16}.

Various nucleosynthetic channels have been proposed to explain the origin of these subclasses.  CEMP-s stars may form in a binary system where the s-process elements are synthesised by and accreted from a companion asymptotic giant branch (AGB) star (e.g. \citealt{Aoki07,Lucatello06,Hansen16}). The CEMP-r process may be operating in neutron star - neutron star mergers (e.g. \citealt{Freiburghaus99, Kasliwal17, Cote18}). CEMP-no stars are thought to be enriched by low metallicity core collapse supernovae \citep[e.g.][]{Frebel08,Nomoto13,Keller14,Choplin17}. 

How to assign an observed CEMP star to the various subclasses is not universally agreed.
\citet{Aoki07} uses the criterion [Ba/Fe]$>1$ to be a CEMP-s as do \citet{Masseron10}  who in addition require [Ba/Fe]$<0$ for a star to be CEMP-no. With, AGB stars, the likely sites of the s-process, as evidenced by many barium enhanced stars having a binary companion which used to be an AGB star \citep{Lucatello06, Hansen16}, the carbon in CEMP-s stars itself could originate from the AGB star. Therefore the absolute carbon abundance by number, A(C)$\equiv \log({\rm N}_{\rm C}/{\rm N}_{\rm H})+12$, might be a good indicator of whether a star is CEMP-s or CEMP-no \citep{Spite13, Bonifacio15, Yoon16}. Using this criterion, \citet{Yoon16} recognised	that CEMP stars with very low [Fe/H] fall into two further subgroups, a subgroup~II with higher A(C) abundance than subgroup~III at the same (low) [Fe/H] (with subgroup~I stars at higher [Fe/H] (see Fig.~1 in \citealt{Yoon16}).

\citet{Chiaki17} suggest that the subgroup~III stars form from gas cooling onto carbon grains while the subgroup-II form via cooling onto silicate grains. Taking a step further, \citet{Hartwig18} argued that the subgroup~II stars were enriched by faint low metallicity  core collapse supernovae (SNe), whilst the subgroup~III stars were enriched by a mix of faint and high mass SNe.

In this {\it Letter} we use the \eagle\ cosmological hydrodynamical simulation described by \citet{Schaye15} to investigate the abundance patterns of CEMP stars. \eagle\ is a cosmological simulation with subgrid parameters calibrated to produce an accurate stellar mass function in the present day Universe, and to produce good agreement with the observed mass metallicity relations for galaxies \citep{Crain15}. We have recently shown that the \eagle\ simulation is also able to produce both CEMP-s and CEMP-no stars \citep{Sharma18} that form in the poorly mixed
	interstellar medium of the first galaxies. Here we investigate whether \eagle\ CEMP stars occur in
	the three subgroups described above, and if so how these originate. This {\it Letter} is organised as follows. In section~\ref{sec_sim}, we describe the subgrid implementation of the \eagle\ simulation. In section~\ref{sec_res} we report our results and describe the physical mechanism responsible for the production of the subgroups of CEMP stars. We summarise our findings in section~\ref{sec_sum}. 
\section{The {\sc eagle} simulation}
\label{sec_sim}
\eagle\ \citep{Schaye15} is a set of cosmological hydrodynamical simulations performed with the smoothed particle hydrodynamic (SPH) code {\sc gadget} \citep{Springel05}. The version of code used for {\sc eagle} includes modification to the hydrodynamics solver to resolve known issues with standard SPH, as well as a set of \lq subgrid\rq\ physics modules to implement the unresolved physics of the interstellar medium, and of feedback from supernovae and accreting supermassive black holes. The simulation is calibrated to reproduce the galaxy stellar mass function, the relation between galaxy mass and galaxy size, and the relation between galaxy mass and black hole mass; see \cite{Crain15} for details. Galaxies are identified in post-processing using the {\sc subfind} algorithm \citep{Springel01, Dolag09}. The data from the simulation are available in a public database  \citep{McAlpine16, Eagle17}.

We briefly summarise the subgrid modules implemented in \eagle\ that are relevant for this study (see \citealt{Schaye15} for full details).  Star formation is implemented as described by \cite{Schaye08}, where a collisional gas particle is stochastically converted to a collisionless star particle above a metallicity-dependent density threshold. The implementation of stellar evolution follows \cite{Wiersma09b}, and takes into account nucleosynthesis and timed enrichment by AGB stars, SNe of type~Ia, and massive stars and their type~II SNe descendants. The yields from the type~II SNe depend on the mass and metallicity of the progenitor star. The simulation tracks 11 elements (H, He, C, N, O, Ne, Mg, Si, S, Ca, Fe) as well as a total metallicity (metal mass) variable for each star and gas particle. In addition, we track the total mass a gas particle received from each of the three enrichment channels ({\em i.e.} AGB, type~Ia  and type~II SNe). Below we use the mass received from the AGB channel, $f_{\rm AGB}$, and from type~II SNe, $f_{\rm SNII}$, defined as
\vspace{-0.25mm}
\begin{equation} 
f_{\rm AGB} = \frac{m_{\rm AGB}}{m_\star}, \qquad f_{\rm SNII} = \frac{m_{\rm SNII}}{m_\star},
\label{eq_frac}
\end{equation}
	where $m_\star$ is the total (initial) mass of the star; the fraction of mass acquired from type~I SNe is always small. These three fractions add up to the fraction of mass in a gas particle that has passed through at least one generation of stars but is not equal to the metallicity (metal mass fraction) because stars expel hydrogen and helium as well, in addition to metals. As a star particle evolves, it spreads the mass lost through stellar evolution to its neighbouring gas particles. Given the $\sim 2\times 10^6{\rm M}_\odot$ mass of each star and gas particle, we cannot hope to resolve the intricate mixing of ejecta of different stellar evolutionary channels. However, we will show that the lack of metal mixing that we find at low metallicity is due to the large-scale dynamics of inflows and outflows of gas in the first galaxies, something that the simulation can resolve.

Energetic feedback from stellar winds and SNe is implemented as a thermal injection of energy. This leads to vigorous outbursts (outflows) of gas from small galaxies, in a series of \lq breathing modes\rq\ where the galaxy loses enriched gas in an outflow, and then gains fresh, mostly primordial gas through cosmological accretion \citep{Sharma18}. Other much higher resolution simulations demonstrate similar behaviour \citep[e.g.][]{Stinson07,Muratov15}. The burstiness creates an environment in which stars can form out of gas that is predominately enriched by either the AGB or type~II SNe channels (but not both). This aspect of early star formation is crucial in understanding the abundances of low metallicity stars in \eagle\ - and potentially of CEMP stars in the Milky Way.
 
Here we use the \eagle\ simulation labelled Recal-L025N0752 in \citet{Schaye15}, that is run in a cubic volume with comoving sides of 25~Mpc. The mass of a dark matter particle is $1.21\times10^6$~M$_\odot$, and the initial mass of a gas particle is $2.26\times10^5$~M$_\odot$. The Plummer equivalent gravitational softening length is taken to be $\epsilon=0.35~{\rm kpc}$ at $z=0$. We select MW-like galaxies from Recal-L025N0752 by selecting haloes of mass $10^{12}\leq M_h\leq 10^{12.5}h^{-1}$M$_\odot$. In those galaxies, we denote a star as CEMP if it has [Fe/H]$\leq-1$ and [C/Fe]$\geq0.7$, following \cite{Yoon16}.

\section{Results}
\subsection{Abundance patterns of CEMP stars in \eagle}
\label{sec_res}
\begin{figure}
\centering
\includegraphics[width=\linewidth,height=14cm]{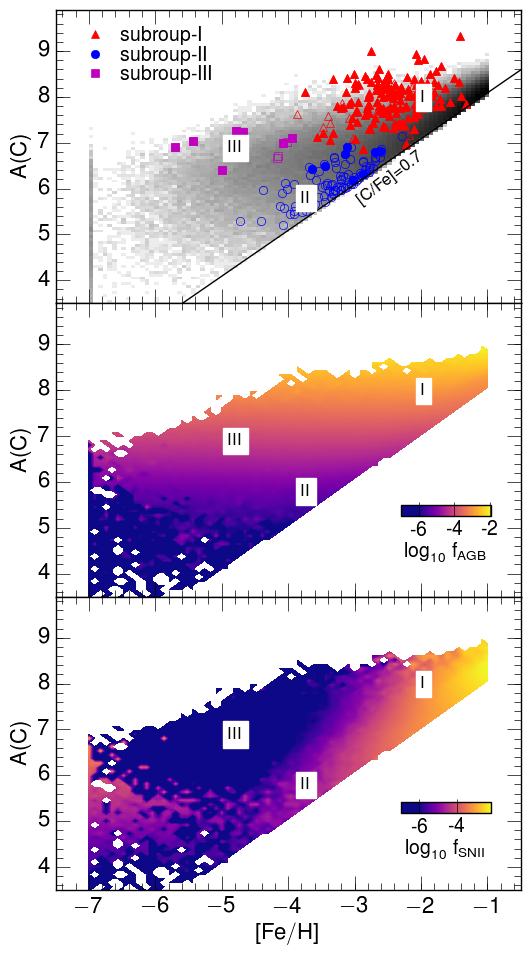}
\caption{{\it Top panel}: the absolute abundance of carbon, A(C), as a function of [Fe/H] for CEMP stars ([Fe/H]$<-1$, [C/Fe]$\geq0.7$ (solid black line)) in \eagle\  is shown as a two dimensional histogram of the number density of CEMP stars (darker a pixel, higher its number density). For comparison we also show the stars compiled by \citet{Yoon16}, who divided them into subgroup~I (red triangles), subgroup~II (blue circles) and subgroup~III (purple squares). Filled symbols correspond to stars with [Ba/Fe]$>0$. {\it Middle panel}: same as top panel, but with the colour coding illustrating the	median value of $f_{\rm AGB}$, the fraction of stellar mass acquired through the AGB channel (Eq.~\ref{eq_frac}) in \eagle. {\it Bottom panel}: same as the middle panel but with colour a measure of the median value of $f_{\rm SNII}$, the fraction of stellar mass acquired through enrichment by type~II SNe (Eq.~\ref{eq_frac}).}
\label{fig_YB_1}
\end{figure}
\begin{figure}
\centering
\includegraphics[width=\linewidth,height=7cm]{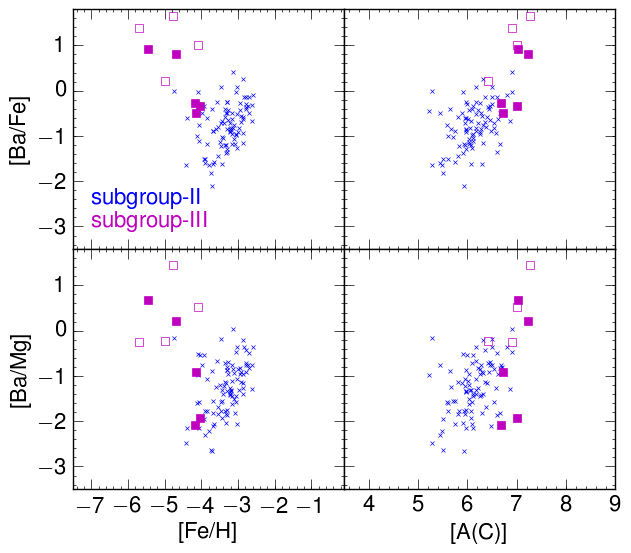}
\caption{Abundance ratios of [Ba/Fe] ({\em bottom row}) and [Ba/Mg] ({\em top row}) versus [Fe/H] ({\em left column}) and A(C) ({\em right column}), for 
	the stars in subgroups~II ({\em blue points}) and III ({\em purple points})
	from Fig.~\ref{fig_YB_1}. The subgroup-III stars with upperlimits on Ba are shown as open squares. }
\label{fig_YBa} 
\end{figure}
\begin{figure}
\centering
\includegraphics[width=\linewidth,height=10cm]{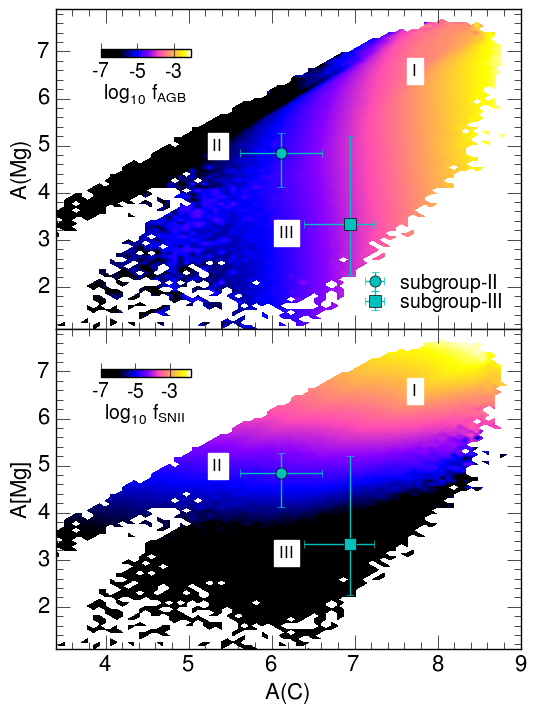}
\caption{Absolute abundance of magnesium, A(Mg), as a function of absolute abundance of carbon, A(C). The {\em top panel} is coloured according to the median $f_{\rm AGB}$ value in \eagle, where $f_{\rm AGB}$ is the mass fraction the gas particle acquired from AGB ejecta before it became a star (Eq.~1); the {\em bottom panel} is coloured with the corresponding mass fraction ejecta by massive stars and their type~II SNe, $f_{\rm SNII}$. Labels I-III indicate the location of the three subgroups of CEMP stars. Median values of A(Mg) and A(C) for CEMP stars of subgroup~II and III from the Milky Way stars observed by \citet{Yoon16} are plotted as a cyan circle and cyan square, respectively, with error bars encompassing the corresponding 10th$\hbox{--}$90th percentiles.}
\label{fig_YMg}
\end{figure}
The subclasses of CEMP stars identified by \cite{Yoon16} are illustrated in the top panel of Fig.~\ref{fig_YB_1}): subgroup~I (red symbols) appears at higher absolute carbon abundance, A(C), and at higher [Fe/H], subgroup~II (blue symbols) is at lower A(C), and subgroup~III (cyan symbols) at lower [Fe/H]. Stars in subgroup-II typically have a lower abundance of carbon when compared to subgroup-III, but higher abundances of Mg and Na \citep{Yoon16}. We have overlaid labels I--III to guide the eye; these are repeated in the lower panels. 
 Various values of [Ba/Fe] are used in the literature to signify AGB enrichment; we use [Ba/Fe]$>0$ and show those stars with filled symbols in the top panel of Fig.~\ref{fig_YB_1}.

The grey shading is a two dimensional histogram of the fraction of Milky Way CEMP stars with given A(C) and [Fe/H] in \eagle: clearly the simulation produces CEMP stars with a large scatter in A(C) at any [Fe/H], accurately covering the range in A(C) as a function of [Fe/H] occupied by observed CEMP stars. The number density of stars decreases with decreasing [Fe/H] in \eagle; nevertheless we find stars that populate the regions designated as subgroup~II and subgroup~III by \cite{Yoon16} but with no obvious gap in number density between the two subgroups.

The origin of the CEMP subgroups in \eagle\ is revealed by colouring the location in the A(C)-[Fe/H] diagram of Fig.~\ref{fig_YB_1} by the fraction of mass acquired from AGB enrichment ($f_{\rm AGB}$, {\em middle panel}) or from type~II SNe enrichment ($f_{\rm SNII}$, {\em bottom panel}): stars in subgroup~II formed from gas predominantly enriched by ejecta from type~II SNe and stars in subgroup~III by ejecta from AGB stars. The presence of gas that is enriched mostly by AGB ejecta may seem surprising at first, given the very early formation time of these low metallicity stars (typically before redshift $z=6$, not shown), but clearly the more massive AGB stars have short enough main sequence progenitor lifetimes to have an impact on the abundance of low-Z stars. Possibly even more surprising is that the natal gas of the subgroup~III stars was not also enriched by type~II SNe, given the even shorter main sequence lifetimes of their progenitors. We will show below that this lack of mixing of ejecta is a direct consequence of the bursty nature of the onset of star formation in the progenitor galaxies.

A key distinction between the ejecta of low mass AGB stars and that of massive stars that undergo core collapse is in their yields of Ba and Mg. Whereas AGB stars produce s-process elements such as Ba but little or no Mg and no Fe, massive stars and their type~II SNe descendants synthesise Mg but no Ba. At low-Z, type~II SNe also expel very little or no Fe due to collapse of the stellar core to a black hole following the supernova explosion (often called \lq fall back\rq, see {\em e.g.} \citealt{Woosley95}). Combined with our claim of the different enrichment channels of subgroups~II and III stars leads us to predict that subgroup~II stars should have low [Ba/Mg] and [Ba/Fe] (consistent with AGB nucleosynthesis), and subgroup~III stars should have high [Ba/Mg] and [Ba/Fe] (consistent with massive stars/core collapse nucleosynthesis). The data plotted in Fig.~\ref{fig_YBa} appear consistent with this interpretation but note that the Ba measurments for some stars are upper limits (open squares).

As also discussed in the Introduction, stars in CEMP subgroup~II formed from gas enriched by AGB stars are also likely to be carbon rich because AGB stars also synthesise carbon, suggesting that the [C/Mg] ratio is also good at distinguishing AGB enrichment from enrichment by type~II SNe \citep{Yoon16,Hartwig18}. This prediction is easily verified in the \eagle\ simulation: the greater the extent stars form from gas that is enriched by AGB ejecta (higher values of $f_{\rm AGB}$), the higher A(C), with little correlation between $f_{\rm AGB}$ and A(Mg) since the Mg in these stars is not acquired from AGB enrichment (top panel of Fig.~\ref{fig_YMg}). As is clear from the labels, stars of subgroup~III are found in the region where the enrichment is mostly by AGB stars, and hence these stars are expected to have low abundances of elements not produced in AGB stars such as for example Mg and Na, consistent with observed abundances of these stars \citep{Yoon16}. Conversely, the bottom panel of Fig.~\ref{fig_YMg} shows that there is a strong correlation between A(Mg) and $f_{\rm SNII}$, the fraction of mass acquired from massive star ejecta that are rich in Mg, but not in A(C) - a consequence of the fact that low-$Z$ type~II SNe produce relatively little carbon. 

By selecting CEMP-no stars in \eagle\ with the criterion A(C)~$<7.1$, and further separating out subgroup~III CEMP-no stars from subgroup~II by the criterion [C/Mg]~$>2.5$, we find that $14$ percent of CEMP-no stars in \eagle\ belong to subgroup-III. This fraction is in reasonable agreement with the observed fraction inferred by \citet{Yoon16}, who find that 12 out of 127 of their CEMP-no stars are in subgroup~III.

\subsection{The origin of abundance patterns of CEMP stars}
\begin{figure}
\centering
\includegraphics[width=\linewidth,height=12cm]{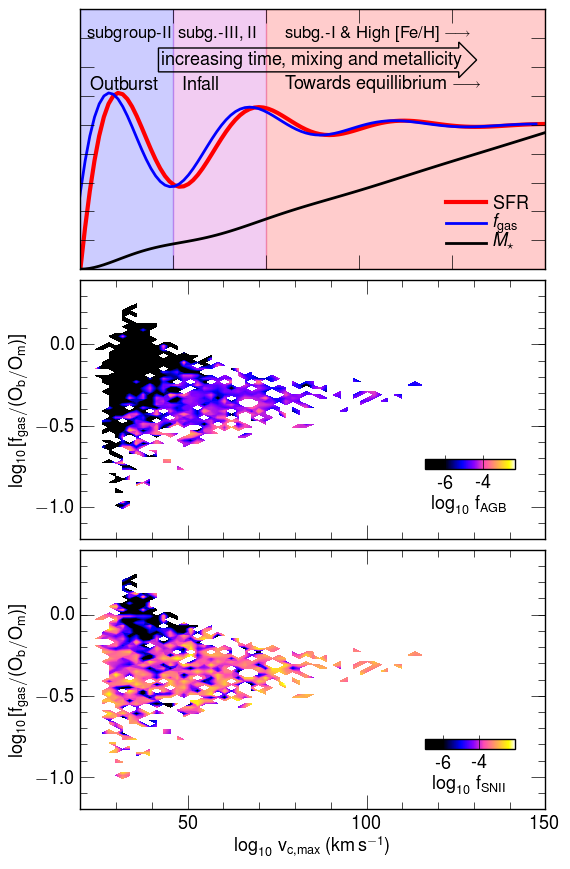}
\caption{{\it Top panel}: a schematic diagram (not to scale) of the onset of star formation in a galaxy, illustrating the large variations in gas fraction that occur before self-regulated star formation ensues in the later stages of formation. The {\em red}, {\em blue} and {\em black} lines are the star formation rate, gas fraction, and stellar mass of the galaxy, respectively, as a function of cosmic time, illustrating the star burst leading to a strong outflow (blue coloured region labelled outburst), followed by cosmological gas accretion re-igniting star formation (purple coloured region labelled infall), and finally self-regulated star formation (coloured red). Annotations subroup~I--~III refer to the epoch at which CEMP stars of subgroups~I, II and III form (see text for details). {\em Middle panel:} gas fraction in units of the cosmic mean baryon fraction, $f_{\rm gas}/(\Omega_m/\Omega_b)$, as a function of the maximum circular velocity, $v_{\rm c,max}$, for \eagle\ galaxies above redshift $z=6$. Galaxies are coloured by the median value of $f_{\rm AGB}$ (Eq.~1) of stars formed within the last 20~Myr. {\em Bottom panel}: same as the upper panel but for $f_{\rm SNII}$. }
\label{fig_gasfrac}
\end{figure}
We have shown that the appearance of characteristic abundance patterns of CEMP stars can be traced back to predominant enrichment of their natal gas clouds by a single nucleosynthetic channel - either operating in AGB stars or in massive stars. In this section we explore the origin of the lack of mixing of the stellar ejecta.

An important clue lies in the way a young galaxy forms stars. Hydrodynamical simulations of galaxy formation find that an infant galaxy goes through one or several {\it breathing} phases, where feedback from massive stars powers a strong outflow which quenches star formation, after which cosmological accretion of primordial gas replenishes the galaxy with gas reigniting star formation \citep[e.g.][]{Stinson07,Muratov15}. A cartoon picture of this scenario is shown in the top panel of Fig.~\ref{fig_gasfrac} (see also Fig.~2 in \citealt{Sharma18}).

The burstiness of star formation in young galaxies strongly influences the abundances of stars that form subsequently \citep[e.g.][]{Sharma18}. During the initial burst (blue shaded region in the top panel of Fig.~\ref{fig_gasfrac}) the low-$Z$ supernovae enrich the gas with magnesium and a little carbon (and with negligible Fe and Ba); the CEMP stars that form out of this gas are of subgroup~II. This initial burst results in a strong outflow that clears most of the gas from the galaxy. As a consequence, the star formation rate decreases and the outflow is quenched. This enables cosmological accretion of mostly primordial gas to replenish the galaxy (purple shaded region) that is then enriched by AGB ejecta of stars that formed in the burst. Since these AGB stars mainly produce carbon and some barium, but little or no magnesium or other heavier elements, CEMP stars that form from this gas are mainly of subgroup~III. As the galaxy matures and its potential well deepens, star formation becomes self-regulating, and star forming gas is more metal rich and better mixed, enabling the formation of subgroup-I CEMP stars.

Evidence that this mechanism is at work in \eagle\ galaxies is provided by the middle and lower panels of Fig.~\ref{fig_gasfrac}, which show the gas fraction, $f_{\rm gas}$, as a function of maximum circular velocity, $v_{\rm c, max}$, of redshift $z>6$ \eagle\ galaxies. When $f_{\rm gas}$ is high, galaxies are in the starburst phase, and CEMP stars of subgroup~II form from gas with high $f_{\rm SNII}$. Conversely, when $f_{\rm gas}$ is low, CEMP stars of subgroup~III form from	gas with a high value of $f_{\rm AGB}$. Finally when the circular velocity is high, the range in gas fraction becomes small because star formation regulates $f_{\rm gas}$, and CEMP stars of subgroup~I form, that are enriched by both the AGB and the massive star nucleosynthetic channels.
\vspace{-7mm} 
\section{Summary and Conclusion}
\label{sec_sum}
Observed carbon enhanced metal poor (CEMP) Milky Way stars can be divided in three different subgroups depending on their absolute carbon abundance at a given value of [Fe/H] \citep{Yoon16}. We have shown that similar abundance patterns are evident in CEMP stars that form in Milky Way-like galaxies of the \eagle\ \citep{Schaye15} cosmological hydrodynamical simulation. In \eagle, these subgroups are a direct result of the bursty nature of star formation in the progenitors of these galaxies, which results in poor mixing of the metal enhanced ejecta of massive stars and type~II SNe compared to ejecta from AGB stars. Stars that form during a burst have characteristics of subgroup~II CEMP stars: high in [Mg/C] with little or no barium. Conversely, stars that form after a burst have characteristics of subgroup~III CEMP stars: higher in C and Ba but low in Mg. As the galaxy grows in mass and its gravitational potential deepens, ejecta of AGB and type~II SNe mix better, resulting in the formation of subgroup~I CEMPs, and eventually higher metallicity stars.

A detailed census of the stars in the Milky Way and its dwarf satellites, and in particular a census of  the metal poor population in these systems, promises to be valuable tool to study the onset of star formation in their progenitors. The variety of abundance patterns in CEMP stars already suggests that star formation was initially very bursty, which bodes well for the detectability of high-$z$ galaxies.
\vspace{-5mm}
\section{Acknowledgements}
We thank Tim Beers for illuminating discussions on the topic of this {\it Letter} during his visit to Durham, and an anonymous referee for constructive remarks. MS was funded by STFC grant ST/P000541/1. This work used the DiRAC Data Centric system at Durham University, operated by the Institute for Computational Cosmology on behalf of the STFC DiRAC HPC Facility (www.dirac.ac.uk). This equipment was funded by BIS National E-infrastructure capital grant ST/K00042X/1, STFC capital grant ST/H008519/1, and STFC DiRAC Operations grant ST/K003267/1 and Durham University. DiRAC is part of the National E-Infrastructure.

\vspace{-5mm}

\bibliographystyle{mnras}
\bibliography{ref_twoCEMPno}

\end{document}